\documentclass{aastex}                 
\usepackage{emulateapj5,apjfonts,epsf}
 
\received{date1}
\accepted{date2}
\journalid{}{}
\articleid{}{}

\journalinfo{{\sc The Astrophysical Journal}, 2002, in press}

\slugcomment{Accepted for publication by The Astrophysical Journal Letters,
September 23, 2002}

\shortauthors{Camilo et al.}
\shorttitle{PSR~J1747$-$2958 and the Mouse}

\begin{document}

%
%

\def\psr{PSR~J1747$-$2958}
\def\pwn{G359.23$-$0.82}
\def\snrA{G359.1$-$0.5}
\def\snrB{G359.0$-$0.9}
\def\chandra{{\em Chandra\/}}
\def\rosat{{\em ROSAT\/}}
\def\bsax{{\em BeppoSAX\/}}

\title{Heartbeat of the Mouse: a young radio pulsar associated with the
axisymmetric nebula G359.23$-$0.82}

\author{F.~Camilo,\altaffilmark{1}
  R.~N.~Manchester,\altaffilmark{2}
  B.~M.~Gaensler,\altaffilmark{3} and
  D.~R.~Lorimer\altaffilmark{4} }
\medskip
\altaffiltext{1}{Columbia Astrophysics Laboratory, Columbia University,
  550 West 120th Street, New York, NY~10027}
\altaffiltext{2}{Australia Telescope National Facility, CSIRO,
  P.O.~Box~76, Epping, NSW~1710, Australia}
\altaffiltext{3}{Harvard-Smithsonian Center for Astrophysics, 60 Garden
  Street, Cambridge, MA~02138}
\altaffiltext{4}{University of Manchester, Jodrell Bank Observatory,
  Macclesfield, Cheshire, SK11~9DL, UK}

\begin{abstract}
We report the discovery of \psr, a radio pulsar with period $P =
98$\,ms and dispersion measure $\mbox{DM} = 101$\,cm$^{-3}$\,pc, in a
deep observation with the Parkes telescope of the axially-symmetric
``Mouse'' radio nebula (\pwn).  Timing measurements of the newly
discovered pulsar reveal a characteristic age $P/2\dot P = 25$\,kyr and
spin-down luminosity $\dot E = 2.5 \times 10^{36}$\,erg\,s$^{-1}$.  The
pulsar (timing) position is consistent with that of the Mouse's
``head''.  The distance derived from the DM, $\approx 2$\,kpc, is
consistent with the Mouse's distance limit from H{\sc i} absorption, $<
5.5$\,kpc.  Also, the X-ray energetics of the Mouse are compatible with
being powered by the pulsar.  Therefore we argue that \psr, moving at
supersonic speed through the local interstellar medium, powers this
unusual non-thermal nebula.  The pulsar is a weak radio source, with
period-averaged flux density at 1374\,MHz of 0.25\,mJy and luminosity
$\sim 1$\,mJy\,kpc$^2$.

\end{abstract}


\section{Introduction}\label{sec:intro} 

The ``Mouse'' (\pwn\footnote{This source previously has been referred
to by several names: G359.23$-$0.82 \cite{yb87}; G359.23$-$0.92
\cite{pk95}; G359.2$-$00.8 ({\sc simbad} database).  In accordance with
IAU rules, we use the original name, even though the indicated position
is not accurate (see Table~\ref{tab:parms}).}; Yusef-Zadeh \& Bally
1987)\nocite{yb87} is among the few known non-thermal radio nebulae
with axial symmetry (see Fig.~\ref{fig:mouse}), consisting of a bright
``head'' and a long ``tail'' that is highly linearly polarized.  It is
also an X-ray source \cite{pk95,smi+99}.  The few other examples known
in this class are manifestations of a pulsar bow shock:  the
relativistic wind of a neutron star confined by ram pressure due to the
supersonic motion of the pulsar through the local interstellar medium
(ISM).  The detailed study of such objects can lead to constraints on
the local ISM density and pulsar velocities, ages, spin and magnetic
field evolution, and winds \cite{cc02,vag+02}, as exemplified by the
study of the ``Duck'' nebula and its pulsar \cite{gf00}.  Also,
relatively few young nearby pulsars are known.  Detecting other such
nearby pulsars, as is likely to be lurking inside the Mouse, is
important for accurately determining pulsar birth rates, beaming
fractions and luminosity distributions (e.g., Brazier \& Johnston
1999)\nocite{bj99}.

The Mouse has therefore been the object of considerable interest since
its discovery.  While its interpretation as a pulsar-powered nebula is
appealing \cite{pk95}, no central engine had been detected in previous
radio pulsation searches.  In this Letter we report the discovery of a
faint young pulsar coincident with the Mouse's head, confirming that
the Mouse is a synchrotron nebula powered by a high velocity neutron
star.

\section{Observations}\label{sec:obs}

The most sensitive previous pulsar search at the position of \pwn\ was
the Parkes Multibeam Pulsar Survey of the inner Galactic plane
\cite{mlc+01}.  The Mouse is only one degree away from the Galactic
center, where background synchrotron emission degrades considerably the
system sensitivity, and we estimate that the flux density limit at the
position of \pwn\ was no better than 0.6\,mJy at a frequency of
1374\,MHz.  Following the recent discovery at Parkes of the very faint
pulsar J1124$-$5916 located in the supernova remnant (SNR) G292.0+1.8
\cite{cmg+02}, we began the deepest searches possible at Parkes of a
number of good young pulsar candidates, including the Mouse.

On 2002 February 1 we searched the head of the Mouse using the center
beam of the multibeam receiver at a central frequency of 1374\,MHz.
The observing setup was identical to that used in the discovery of
PSR~J1124$-$5916, with total-power signals from 96 frequency channels,
spanning a total band of 288\,MHz for each of two polarizations,
sampled at 1\,ms intervals and recorded to magnetic tape for off-line
analysis. The total observation time was 9.4\,hr.  The data were
reduced in standard fashion using an FFT-based code (see Lorimer et
al.~2000\nocite{lkm+00} for details) to search for periodic signals in
de-dispersed time series with trial dispersion measures (DMs) in the
range 0--8800\,cm$^{-3}$\,pc. For further details, see Camilo et
al.~(2002b)\nocite{cmg+02}.  A clear periodic and dispersed signal with
period $P \approx 98.8$\,ms was detected with maximum signal-to-noise
ratio of 15.1 at $\mbox{DM} \approx 105$\,cm$^{-3}$\,pc.

\begin{figure}
\plotone{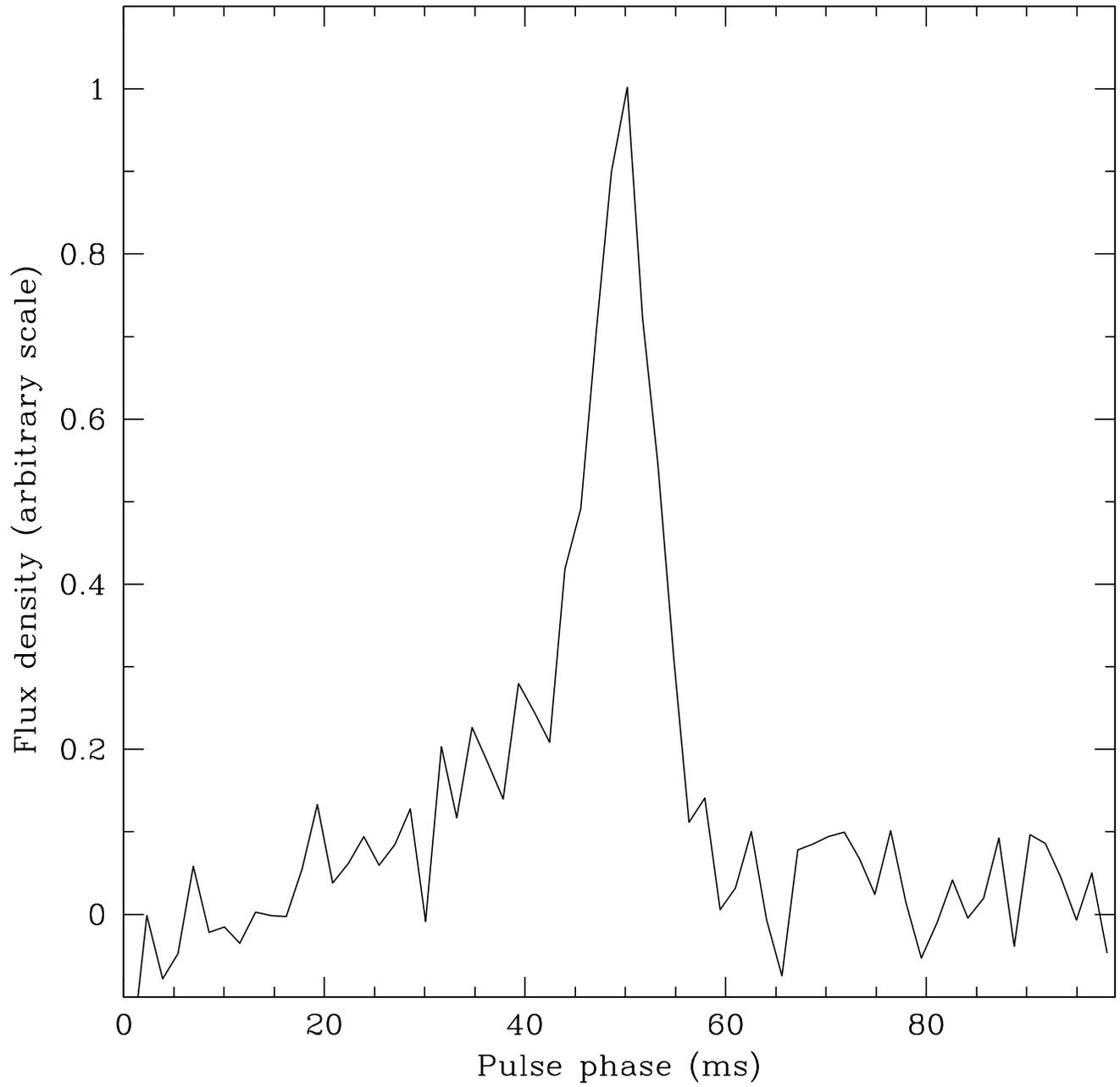}
\caption{\label{fig:prof} 
Mean pulse profile of \psr\ at 1374\,MHz, based on 56\,hr of Parkes
observations. }
\end{figure}

We confirmed the pulsar with a 3\,hr observation on February 3 and
thereafter began similar regular timing observations, thus far having
obtained 17 times-of-arrival (TOAs) spanning 6 months.  We have used
the {\sc tempo}\footnote{See http://pulsar.princeton.edu/tempo.} timing
software and the TOAs to derive the ephemeris listed in
Table~\ref{tab:parms}.  Based on the coordinates of this pulsar, we
designate it as \psr.  The observed flux density of the pulsar is
somewhat variable, in a manner consistent with interstellar
scintillation.  We formed a mean pulse profile, shown in
Figure~\ref{fig:prof}, by averaging all available observations. This
shows that the pulsed emission covers about 50\% of the pulse period,
rising slowly to the peak and falling sharply after it. This profile
was calibrated (for details, see Manchester et
al.~2001)\nocite{mlc+01}, taking into account the measured sky
brightness temperature at the Mouse position (18\,K), to derive a flux
density of $S = 0.25 \pm 0.03$\,mJy.  The radio luminosity $S d^2$ for
a distance $d \approx 2$\,kpc (see \S~\ref{sec:disc}) is $\sim
1$\,mJy\,kpc$^2$.  This is a very low value, but is similar to those of
four young and energetic pulsars discovered recently (Halpern et
al.~2001; Camilo et al.~2002a--c)\nocite{hcg+01,cmg+02,csl+02,clb+02},
highlighting that such pulsars can be very faint radio sources.

On 2002 May 15 we attempted, unsuccessfully, to measure the position of
the pulsar independent of pulsar timing by performing pulsar-gated
radio imaging with the Australia Telescope Compact Array (ATCA).  We
observed the field of the Mouse in the 6A array configuration
simultaneously with two $32 \times 4$-MHz bandwidths centered at
frequencies of 1.4 and 1.7\,GHz for an effective on-source integration
time of about 11\,hr. PKS~1934$-$638 was the primary flux density
calibrator and PKS~1748$-$253 was the secondary phase calibrator.
While pointing at \psr, data for each baseline were integrated into 32
pulse phase bins coherently with the topocentric period of the pulsar.
We also made a number of 5\,min integrations on a strong pulsar,
PSR~B1641$-$45.  The data were analyzed in standard fashion with the
{\sc miriad}\footnote{See
http://www.atnf.csiro.au/computing/software/miriad.} software (see,
e.g., Stappers, Gaensler, \& Johnston 1999)\nocite{sbj99} eventually
yielding ``on-pulse $-$ off-pulse'' images.  While PSR~B1641$-$45 was
detected easily, \psr\ was not, with an upper limit on the flux density
averaged over $\sim 0.1$ of the period of about 2\,mJy. This is
comparable to the expected peak flux density of the pulsar.


In Figure~\ref{fig:mouse} we display two radio images of the Mouse.  In
the left panel we show a $42' \times 42'$ field taken from the MOST
Galactic Center Survey (MGCS; Gray 1994)\nocite{gra94c}, with the
field-of-view of our Parkes observation overlaid.  In the right panel
we show a higher resolution image of the head of the Mouse, expanded by
a factor of 30 with respect to the left panel.  This was obtained from
an observation with the Very Large Array (VLA) in its A configuration
on 1999 October 8 at a frequency of 8.4\,GHz, using a bandwidth of
100\,MHz with an on-source integration time of 3\,hr, and analyzed in
standard fashion.

\section{Discussion}\label{sec:disc}

The $P$ and $\dot P$ measured for \psr\ imply a relatively large
spin-down luminosity $\dot E = 4\pi^2 I \dot P /P^3 = 2.5 \times
10^{36}$\,erg\,s$^{-1}$ (where the neutron star moment of inertia $I
\equiv 10^{45}$\,g\,cm$^2$), small characteristic age $\tau_c = P/2
\dot P = 25$\,kyr, and surface magnetic dipole field strength $B =
3.2\times10^{19} (P \dot P)^{1/2} = 2.5 \times 10^{12}$\,G.  These
parameters place \psr\ in the group of $\approx 20$ ``Vela-like''
pulsars now known (those with $\dot E \ga 10^{36}$\,erg\,s$^{-1}$ and
$10 \la \tau_c \la 100$\,kyr).  In the remainder of this section we
discuss the evidence linking \psr\ with the Mouse.

\begin{figure}
\plotone{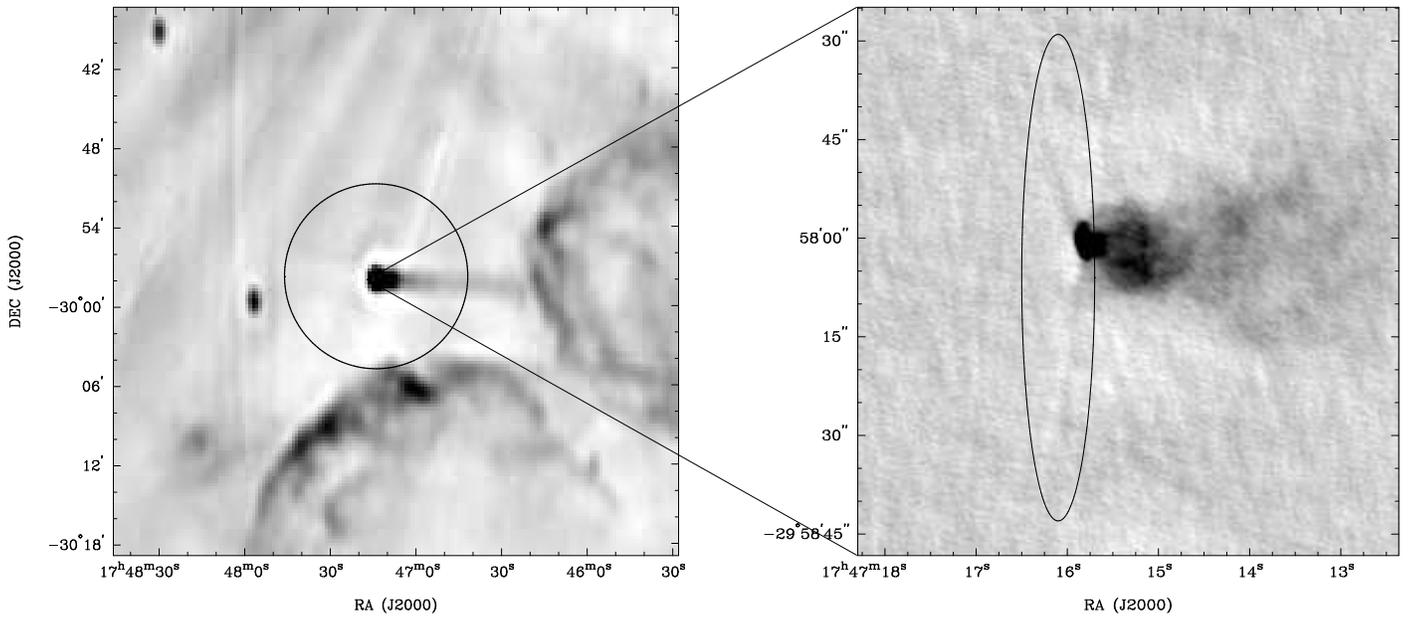}
\caption{\label{fig:mouse} 
Radio images of the Mouse.  {\em Left:\/} Large scale view at 0.8\,GHz
from the MGCS with an angular resolution of $43''$, showing the Mouse's
bright head and long tail, and the unrelated (Uchida, Morris, \&
Yusef-Zadeh 1992)\protect\nocite{umy92} SNRs \snrA\ (west) and
\snrB\ (south).  {\em Right:\/} Detailed view of the Mouse's head from
VLA observations at 8.4\,GHz with resolution of $0\farcs9 \times
0\farcs4$ and rms sensitivity of 20\,$\mu$Jy\,beam$^{-1}$.  The head is
clearly resolved in this image.  At the estimated distance of 2\,kpc
(see \S~\ref{sec:disc}), $0\farcs5$ corresponds to a linear size of
1000\,AU.  Notice the relatively faint ``wake'' of nearly triangular
shape, with apparent (semi) opening angle $\sim 30\arcdeg$.  The
ellipse denotes the current positional uncertainty of \psr, determined
from timing observations (Table~\ref{tab:parms}). }
\end{figure}

With the present positional accuracy (Table~\ref{tab:parms} and
Fig.~\ref{fig:mouse}), the offset between the pulsar's timing position
and that of the head of the Mouse seen in the right panel of
Figure~\ref{fig:mouse} is $7'' \pm 37''$, and the area of the error
ellipse is $5 \times 10^{-5}\,\deg^2$.  With approximately 1000 pulsars
known in an area $\approx 1000\,\deg^2$ along the inner Galactic plane
($260\arcdeg \la l \la 100\arcdeg$; $|b|\la 2\fdg5$), the probability
of finding one by chance this close to the Mouse's head is about $5
\times 10^{-5}$.  We therefore regard the positional match of both
sources as highly suggestive of an association.  Eventually a more
precise position for the pulsar will be obtained from timing, and
possibly from \chandra\ observations.

We now consider distance indicators.  The measured DM, together with
the Cordes \& Lazio (2002)\nocite{cl02} model for the Galactic
distribution of free electrons, implies a pulsar distance of 2\,kpc
(the older model of Taylor \& Cordes 1993\nocite{tc93} yields $2.1 \la
d \la 2.8$\,kpc).  The distance to the Mouse has been investigated
using H{\sc i} absorption measurements.  Owing to the lack of
absorption against a ring located 3\,kpc from the Galactic center,
Uchida et al.~(1992)\nocite{umy92} infer that the Mouse is located at
$<5.5$\,kpc from the Sun.  This is consistent with the pulsar distance
determination, and hereafter we consider both objects to be located at
$\approx 2$\,kpc and parametrize the distance in terms of $d_2 =
d/2$\,kpc.

The head of the Mouse has been detected in X-rays, although with
limited statistics \cite{pk95} and angular resolution \cite{smi+99}.
Sidoli et al. model the source with a power-law spectrum having
unabsorbed 2--10\,keV flux $\approx 3 \times
10^{-11}$\,erg\,cm$^{-2}$\,s$^{-1}$.  Assuming this flux to be
isotropic implies a luminosity $L_X \sim 1.4 \times
10^{34}\,d_2^2$\,erg\,s$^{-1}$, or an efficiency for conversion of
spin-down luminosity into X-ray emission of $L_X/\dot E \sim
0.005\,d_2^2$.  This is apparently a factor of $\sim 4$ larger than the
comparable efficiency in the PSR~B1757$-$24/Duck pulsar wind nebula
\cite{kggl01}, a system displaying a bow-shock morphology \cite{gf00}
and seeming in many respects similar to the Mouse/\psr, including a
pulsar with comparable spin parameters.  Only X-ray observations with
higher sensitivity and resolution will settle the issue, but the
presently available data do indicate that the X-ray emission observed
from the direction of the Mouse's head is certainly compatible with an
origin in this system, located at a distance of $\approx 2$\,kpc.  


Given the positional coincidence, consistency in distance indicators,
and energetics compatible with a common source, we regard the
\psr/Mouse association as secure.  The morphology of the Mouse (bright
head, coincident with \psr, trailed to the west by a $12'$-long
cometary tail; Fig.~\ref{fig:mouse}) suggests fast motion of the pulsar
through the ambient ISM.  The tail (length $L \approx 7\,d_2$\,pc)
presumably results from synchrotron radiation produced by the pulsar
relativistic wind in the nebular magnetic field.  For a typical field
of tens of $\mu$G, the lifetime of the radiating particles is $\sim
10^6$\,yr.  The non-thermal spectrum steepens away from the pulsar
location likely due to synchrotron losses \cite{yb87}.  Much can be
learned about the pulsar and the local ISM through a detailed study of
the Mouse's head and tail, as we now outline.

As the pulsar moves supersonically through the ambient medium producing
a bow shock, the morphology of the Mouse's head is expected to be
shaped by ram-pressure balance between the pulsar relativistic wind and
the local ISM.  In particular, the standoff radius of the shock, $R_0$,
for a neutron star moving with velocity $V$ through a medium of density
$\rho$, is determined from $\rho V^2 = \dot E/(4 \pi c R_0^2)$ (see,
e.g., Chatterjee \& Cordes 2002)\nocite{cc02}.  We assume that the
pulsar wind is radiated isotropically, and hereafter neglect projection
effects.  We have not measured the standoff angle $\theta_0 = R_0/d$,
but infer a crude estimate from Figure~\ref{fig:mouse} as follows.  The
width of the Mouse's head in the presumed direction of motion (west to
east) is about $1\farcs6$.  It is likely that the pulsar lies within
this region of intense synchrotron emission and that the apex of the
bow shock lies just outside \cite{buc02}.  Thus we estimate $\theta_0
\sim 1''$, parametrized as $\theta_1 = \theta_0/1''$.  With $\rho =
1.37\,m_H n$, where $n$ is the medium's number density, $m_H$ is the
mass of the H atom, and the numerical factor derives from assuming a
cosmic abundance of He in the ISM \cite{cc02}, we obtain $V \sim
570/(d_2 \theta_1 n^{1/2})$\,km\,s$^{-1}$, where we have used the $\dot
E$ measured for \psr.  For a reasonable pulsar velocity ($1000 \ga V
\ga 100$\,km\,s$^{-1}$; e.g., Lyne \& Lorimer 1994)\nocite{ll94}, this
implies an ISM of high density, $0.3 \la n \la 30$\,cm$^{-3}$.  In turn
this suggests that the pulsar is moving (rapidly) through a warm or
cold phase of the ISM (e.g., Heiles 2001)\nocite{hei01}, with attendant
small sound speed $C$, and hence that its Mach number should be high,
$V/C \gg 10$.  We now extend our attention to the Mouse's tail.

We can obtain a crude estimate of the pulsar's age $\tau$ by
considering that it has traveled the observed length of the tail in its
lifetime, $\tau = L/V \approx 12\,d_2/V_{570}$\,kyr, where $V =
570\,V_{570}$\,km\,s$^{-1}$.  The age may differ significantly from
this if only part of the tail has been detected or if the tail is
instead caused by a relatively recent pulsar backflow (e.g., Kaspi et
al.~2001)\nocite{kggl01}.  The actual age compares to the pulsar
characteristic age of $\tau_c = 25$\,kyr.  These are related by $\tau =
2 \tau_c [ 1 - (P_0/P)^{n-1} ]/(n-1)$ under the assumption of constant
magnetic moment, where $P_0$ is the initial period and $n$ is the
braking index of rotation \cite{mt77}.  A number of factors may cause
$\tau$ to be larger, or smaller, than $\tau_c$.  $P_0 \ll P$, and $n =
3$, appropriate to magnetic dipole braking, are usually assumed.
However for some pulsars $n < 3$ (Mereghetti et al.~2002\nocite{mbbi02}
and references therein).  Also, $P_0$ may be large, up to perhaps
$\approx 90$\,ms (e.g., Camilo et al.~2002b)\nocite{cmg+02}.  In light
of these issues, we regard the pulsar's characteristic age as
approximately consistent with that inferred from the length of the
tail, confirming this source as a relatively young pulsar.

In summary, while it appears that the nature of the source powering the
Mouse has finally been uncovered with the discovery of \psr, much
remains to be understood about this fascinating object, requiring
further observational efforts.  A key measurement to be made is the
proper motion of the pulsar, $\mu = 60\,V_{570}/d_2$\,mas\,yr$^{-1}$.
This may be possible through \chandra\ observations and, depending on
the amount of ``timing noise'' present in the neutron star, via radio
timing.  The pulsar--bow-shock standoff distance is also of great
interest.  Finally, further study of the Mouse's tail, independent
characterization of the local ISM, and any improvement to the distance
estimate, would be most useful.

\acknowledgments

We thank Andy Faulkner, Andrea Possenti, John Reynolds, and John
Sarkissian for help with Parkes observations.  The Parkes and ATCA
radio telescopes are part of the Australia Telescope which is funded by
the Commonwealth of Australia for operation as a National Facility
managed by CSIRO.  We thank Dale Frail for his involvement in the VLA
observations.  The National Radio Astronomy Observatory is a facility
of the National Science Foundation, operated under cooperative
agreement by Associated Universities, Inc.  FC acknowledges support
from NASA grant NAG~5-9950 and a travel grant from the National Radio
Astronomy Observatory.  DRL is a University Research Fellow funded by
the Royal Society.

\begin{deluxetable}{ll}
\tablecaption{\label{tab:parms}Parameters of \psr\ }
\tablecolumns{2}
\tablewidth{0pc}
\tablehead{
\colhead{Parameter}   &
\colhead{Value}     \\}
\startdata
R.A. (J2000) [pulsar]\tablenotemark{a}\dotfill&$17^{\rm h}47^{\rm m}16\fs1(4)$\\
R.A. (J2000) [nebula]\tablenotemark{b}\dotfill&$17^{\rm h}47^{\rm m}15\fs8(1)$\\
Decl. (J2000) [pulsar]\tablenotemark{a}\dotfill & $-29\arcdeg58'07''(37)$     \\
Decl. (J2000) [nebula]\tablenotemark{b}\dotfill & $-29\arcdeg58'00''(2)$      \\
Galactic coordinates [nebula]\tablenotemark{b}\dotfill & $359\fdg305$, 
                                                         $-0\fdg841$          \\
Period, $P$ (ms)\dotfill                       & 98.81275773(5)               \\
Period derivative, $\dot P$\dotfill            & $6.136(5)\times10^{-14}$     \\
Epoch (MJD [TDB])\dotfill                      & 52383.0                      \\
Dispersion measure, DM (cm$^{-3}$\,pc)\dotfill & $101.5(16)$                  \\
Flux density at 1374\,MHz (mJy)\dotfill        & $0.25 \pm 0.03$              \\
Pulse FWHM at 1374\,MHz (ms)\dotfill           & $7 \pm 1$                    \\
Distance, $d$ (kpc)\dotfill                    & $\approx 2$                  \\
Derived parameters:                                      &                    \\
~~Characteristic age, $\tau_c$ (kyr)\dotfill             & 25.5               \\
~~Spin-down luminosity, $\dot E$ (erg\,s$^{-1}$)\dotfill & $2.5\times10^{36}$ \\
~~Magnetic field strength, $B$ (G)\dotfill               & $2.5\times10^{12}$ \\
~~Luminosity at 1400\,MHz (mJy\,kpc$^2$)\dotfill         & $\sim 1$           \\
\enddata
\tablecomments{Numbers in parentheses represent 1\,$\sigma$
uncertainties in the last digits quoted and, except as noted, are equal
to twice the formal errors determined with {\sc tempo}. }
\tablenotetext{a}{Derived from timing measurements.  Uncertainty in
Decl. is much larger than in R.A. due to the low ecliptic latitude
($\beta = -6\arcdeg$) of the pulsar. }
\tablenotetext{b}{Derived from VLA measurements (see
Fig.~\ref{fig:mouse}).  Position (uncertainty) is that of center
(extent) of head of the Mouse, an ellipse oriented at P.A. $\approx
0\arcdeg$.  This position is coincident with that of the X-ray source
detected by \rosat\ with $\approx 8''$ uncertainty \cite{pk95}. }
\end{deluxetable}

\clearpage

\clearpage

\end{document}